\documentclass[twocolumn,showpacs,amsmath,amssymb,prb]{revtex4}
\usepackage{times}
\usepackage{graphicx}
\usepackage{wasysym} 

\usepackage{dcolumn}
\usepackage{bm}

\begin{document}

\title{Levitation of quantum Hall critical states in a lattice model with 
spatially correlated disorder}

\author{Th.\@ Koschny and L. Schweitzer}
\affiliation{Physikalisch-Technische Bundesanstalt, Bundesallee 100, 38116
Braunschweig, Germany}

\date{\today}

\begin{abstract}
The fate of the current carrying states of a quantum Hall system is 
considered in the situation when the disorder strength is increased
and the transition from the quantum Hall liquid to the Hall insulator
takes place. We investigate a two-dimensional lattice model with 
spatially correlated disorder potentials and calculate the density 
of states and the localization length either by using a recursive 
Green function method or by direct diagonalization in connection 
with the procedure of level statistics.
From the knowledge of the energy and disorder dependence of the 
localization length and the density of states (DOS) of the corresponding
Landau bands, the movement of the current carrying states in the 
disorder--energy and disorder--filling-factor plane can be traced 
by tuning the disorder strength.

We show results for all sub-bands, particularly 
the traces of the Chern and anti-Chern states as well as the peak
positions of the DOS. 
For small disorder strength $W$ we recover the well known weak
levitation of the critical states, but we also reveal, for larger $W$, 
the strong levitation of these states across the Landau gaps without
merging. 
We find the behavior to be similar for exponentially, Gaussian, and 
Lorentzian correlated disorder potentials. 
Our study resolves the discrepancies of previously published 
work in demonstrating the conflicting results to be only special
cases of a general lattice model with spatially correlated disorder 
potentials.

To test whether the mixing between consecutive Landau bands is the 
origin of the observed floating, we truncate the Hilbert space
of our model Hamiltonian and calculate the behavior of the current
carrying states under these restricted conditions. 

\end{abstract}


\maketitle

\renewcommand{\baselinestretch}{0.833}

\section{Introduction}
A quantum Hall liquid system is believed to undergo a transition to a Hall 
insulator when the strength of the magnetic flux density $B$ is turned
to zero. \cite{JJWH93,SKD93,Wea94,Hea94a,GJJ95}  
This behavior was suggested by Khmelnitskii 
\cite{Khm84} and Laughlin \cite{Lau84} and explained in terms of a levitation 
scenario where, with decreasing $B$, the current carrying states float
up in energy moving one by one across the Fermi level. Correspondingly,
each time the Hall conductivity $\sigma_{xy}$ drops by an amount $e^2/h$, 
until the last current carrying state gets depopulated and $\sigma_{xy}\to 0$. 
This notion provided the basic ingredient for the construction of a global
phase diagram for the quantum Hall effect. \cite{KLZ92}
In the quantum Hall liquid both the longitudinal conductivity
$\sigma_{xx}$ and resistivity $\rho_{xx}$ get unmeasurable small in 
the limit temperature $T\to 0$, whereas the Hall insulator is characterized 
by a diverging resistivity $\rho_{xx}\to \infty$ while $\sigma_{xx}\to 0$.

However, some of the experimental results are not in accordance 
with the floating up picture. For example, direct transitions to the
Hall insulator from higher ($\nu>2$) quantum Hall plateaus have been 
reported \cite{STC95,LCSL98,Hea00}
as well as a saturation of the Landau level shift as $B\to 0$ for high 
quality p-GaAs samples. \cite{DJS98,HaN99} 
Recently, Yasin and co-workers \cite{Yea02} have claimed that all
the conflicting data published so far for different materials 
can be reconciled by a modified global phase diagram. 
They also find that the current carrying states continously float 
up in energy as $B\to 0$, independent of the appearance of an apparent
`metal'-insulator transition at $B=0$.

On the theoretical side, the situation is still unclear. In particular, 
the microscopic origin of the floating remains obscure because the
possible mechanisms put forward hitherto are only able to account for
a very weak energetical shift, but not for 
the floating of the current carrying states across the Landau gap. 
\cite{SR95,GR96,HY97,Fog98} Therefore, numerical investigations were
carried out in order to get some hint for the underlying physical
process that causes the critical states to move upward in energy.
Instead of studying the fate of the current carrying states when the
magnetic field is decreased one can also look at the influence of an
increasing disorder while $B$ is fixed. 
\cite{LXN96,XLSN96,YB96,SW97,SW98,PBS98,YB99,SW00b,MIH00,PS01} 
In both cases, the Landau bands eventually start to overlap and the 
matrix elements connecting disorder broadened neighboring Landau
bands become important. 

The results of the numerical work for a disordered two-dimensional 
lattice model can be summarized as follows: There is a genuine 
floating of the critical states to higher energies.
\cite{YB96,PBS98,YB99,PS01,KPS01,KS02,PS02a} This floating is, however,  
not easy to observe in systems with uncorrelated random disorder 
potentials due to a peculiar annihilation mechanism inherent in
the lattice model. In this model the Chern states, which correspond to 
the critical electronic states in each Landau band that are responsible 
for the integer quantized Hall conductivity, get neutralized by the so 
called anti-Chern states originating from the center of the tight-binding 
band. They move outwards in energy almost up to the band edges when the 
disorder strength is increased.
Using spatially correlated random disorder potentials instead, it could
be shown that the effect of the anti-Chern states is reduced so that the
expected levitation scenario and the floating of current carrying
states across the Landau gap is seen also in the lattice model. 
\cite{KPS01,KS02} These results also imply that, contrary 
to other claims, \cite{SW98,SW00b} the 
direct transitions to the Hall insulator occurring within the 
lattice model with uncorrelated disorder potentials can of course not 
be held responsible for the direct transitions seemingly observed 
in experiments. \cite{Hea00} Here, finite size effects and limited 
resolution due to finite temperatures are the most probable causes 
\cite{Huc00} that account for the reported behavior. 

The major aims of our work are fourfold: 
\textit{i}) We have to check the universality of our previous results 
which were obtained for a special model of exponentially correlated 
disorder potentials. Therefore, we study in the present work the 
influence of Gaussian and, to a lesser extent, Lorentzian like correlated 
disorder potentials in some detail. In particular, the results of the 
Gaussian disorder model will be useful when comparing with available 
analytical \cite{HY97,Fog98} and recent numerical work where a 'universal 
quantitative relation' for the levitation of extended states has been 
proposed \cite{PS02a,PS02b}. 
Also, the correlated Gaussian model may likely become of greater
importance to future analytical work on this subject because of the 
calculational advantage of Gaussian distributions compared to the 
exponential case.

\textit{ii}) In our previous publication \cite{KPS01} the energetical 
shift of the current carrying states was obtained as a function of disorder
strength and potential correlation length. However, the observed shift 
comprises the broadening effect of the total tight binding band and the 
floating up of the critical states in energy relative to the position of 
the density of states peak. 
To distinguish between these individual contributions and to address the 
question which of the possible definitions for the energetical shift 
\cite{LXN96,YB96,PS02a} are both reasonable and practicable, we have 
to calculate the density of states too. In addition we determine the 
levitation of the critical states as a function of filling factor that 
can be compared with experiments more easily. 

\textit{iii}) The general functional dependence of the levitation of the
critical energy on disorder strength and magnetic field is unknown. Besides
the conjectures independently suggested by Khmelnitskii \cite{Khm84} and 
Laughlin \cite{Lau84} there
exits only some approximate analytical work for the weak levitation regime  
\cite{SR95,GR96,HY97,Fog98}. However, even within this limited range the 
various results partially contradict each other. 
Hence, for the advancement of this unsolved problem it seems adjuvant 
to numerically establish some empirical relations between disorder 
strength $W$, correlation length $\eta$, broadening of Landau levels 
$\mathit{\Gamma}$ and the floating up in energy of the critical states.
These relations, obtained for a well defined model, may then serve as a 
starting point for contriving much simpler models and later on as 
a benchmark for checking the outcome of subsequent analytical work. 

\textit{iv}) To provide a clue for a necessary microscopic explanation 
of the levitation scenario, we truncate the disordered lattice model 
and investigate how the floating up of critical states changes in the 
reduced model containing only one or two Landau bands. This approach 
simultaneously provides a new method for the scrutiny of the relevant 
matrix elements. This work is already in progress.

\section{Model and method}
The model consists of non-interacting electrons moving in the $x$-$y$-plane 
in the presence of disorder and a strong perpendicular magnetic field.
This situation can be described by a Hamiltonian defined on the sites
of a 2d lattice of size $M\times L$ with lattice constant $a$.
Choosing the vector potential in the Landau gauge, 
$\mathbf{A}\equiv(0,-Bx,0)$ with
$(\nabla\times \mathbf{A})_{z}=B$, we have
\begin{eqnarray}
\lefteqn{(H\psi)(x,y) \ =\ } & & \nonumber \\
& & w(x,y)\,\psi(x,y) + V\, \big[ \psi(x+a,y)+\psi(x-a,y)\, \nonumber \\
& & \quad + \exp(-i2\pi\alpha_{B}x/a)\,\psi(x,y+a) \nonumber \\
& & \quad + \exp(i2\pi\alpha_{B}x/a)\,\psi(x,y-a) \big],
\end{eqnarray}
where the magnetic field is chosen to be commensurate with the lattice 
size and 
$\alpha_{B}=a^{2}eB/h$ denotes the strength of $B$ expressed as the number
of flux quanta $h/e$ per plaquette $a^2$. We mostly use $\alpha_{B}=1/8$ 
in the present work and set the units of energy and length to $V=1$ and 
$a=1$, respectively. The on-site disorder potentials $w(x,y)$ are either 
uncorrelated or spatially correlated random numbers with probability
density distributions as described in the next section.

The density of states $\rho(E,W,\eta)$
has been calculated by direct diagonalization of
square systems of linear size $M=64$ applying periodic
boundary conditions in both directions. For the calculation of the 
localization length we utilize a recursive Green function method
\cite{MK81,MK83} which allows to determine the transmission through very
long ($L \sim 10^{7}$) quasi-one-dimensional systems of width $M$
in the range $48\le M \le 192$ from which the energetical position of
the extended states can be extracted using a finite size scaling procedure. 

The inverse localization length as a function of energy, disorder strength,
and correlation length
is defined as the exponential decay of the trace over 
the modulus of a sub-matrix of the one-electron Green function
\begin{equation}
\lambda_{M}^{-1}(E,W,\eta)=-\lim_{L\to\infty}\frac{1}{2L}
\ln(\textnormal{Tr}\,|G_{1L}^{+}|^2).
\end{equation}
$G_{1L}^{+}$ is the $M\!\times\!M$ sub-matrix of the Green function 
$G(E)^{\pm}=(E-H\pm i\epsilon)^{-1}$  
acting in the subspace of the columns $1$, $L$ on the lattice. 
For the calculation of $\lambda_{M}^{-1}(E,W,\eta)$ periodic boundary 
conditions are applied only in the
$y$-direction (width of the system).

In the case of the projected disordered Harper model we used 
square systems of linear size $48\le M\le 96$ and applied the method
of level statistics to determine the energetical position of the
current carrying states. This information is contained in the 
probability density distribution of the energy spacings of neighboring 
eigenvalues. Previously, this powerful method has been successfully 
applied to the study of critical properties at certain metal-insulator 
transitions. \cite{Sea93,ZK95,BSZK96,KBS98,PS02} We have checked that
the results calculated by this method 
agree with the ones obtained by the recursive Green function method. 
However, with comparable amount of numerical effort, the energetical 
resolution achieved is better in the Green function technique.

\section{Correlated disorder model}
For the generation of the random disorder potential an algorithm is required 
which computes spatially correlated random numbers with a known 
correlation function and an arbitrary correlation length, continuously
adjustable from zero (the uncorrelated case) to finite values.
To be used efficiently in conjunction with the recursive Green function 
method which operates on narrow but very long stripes, 
the random potential has to be computed ``on the fly'', i.e., 
without keeping the whole random number field in memory.
We generated a correlated random potential on the lattice $\mathcal{G}$ 
by averaging over uncorrelated random numbers $\varepsilon({\bf m})$ 
associated with each 
lattice point using a suitable weighting function $f$.
The new correlated random numbers are defined as
\begin{equation}
\xi({\bf m}) := 
  \frac{1}{N} \sum\limits_{{\bf m}'\in\mathcal{G}}^{}\, f({\bf m}-{\bf m}')\,
    \varepsilon({\bf m}')
\end{equation}
with ${\bf m}=(m_{x},m_{y})\in\mathcal{G}$. The disorder potential is 
$w({\bf m})=W\,\xi({\bf m})$, where $W$ is the disorder strength 
and the uncorrelated random numbers $\varepsilon({\bf m})$ are 
uniformly distributed over $[-1,1]$.
The exact normalization factor $N=\sum_{{\bf m}\in\mathcal{G}}\, f({\bf m})$ 
allows a comparison of different disorder models and correlation lengths
via the homogeneous distribution function 
$p_{\xi({\bf m})}^{}(z)$ (see below) and the second moment 
of the random potential. For long-range correlations
($\eta >1$), $N$ approaches the continuum limit ($\pi\eta^2$ for
Gaussian correlations), but
deviates considerably for small $\eta$, e.g., $\eta \le 0.3$.  

It is expedient to choose $f$ isotrop, $f({\bf m})=f(|{\bf m}|)$ 
and vanishing outside some finite circle $|{\bf m}|<R$.
This cutoff $R$ makes the generation of the $\xi$ local as
we only need to keep a window corresponding to an area  $2R$ times 
the lattice width in memory.
The $\xi({\bf m})$ have a local distribution of realizations $z\in
[-1,1]$ that, with increasing $\eta$, transforms from a box shape into a 
Gaussian. The form of the distribution function can well 
be approximated by
\begin{eqnarray}
   p_{\xi({\bf m})}^{}(z)
  &\approx&
  \frac{N}{4f(0)}\
  \Bigg\{
    \mathrm{erf}
    \Bigg(
      \frac{f(0)N^{-1}-z}
           {\sqrt{\frac{2}{3}\big(g_{2}^{}[f]-f(0)^2N^{-2}\big)}}
    \Bigg) 
    \nonumber \\
    & &    \ +\ 
    \mathrm{erf}
    \Big( z\mapsto -z \Big)
  \Bigg\}
\end{eqnarray}
with a second moment
$ <\xi({\bf m})^2\!>_{\xi}^{} = g_{2}[f]/3 $ and
$g_{n}[f]=N^{-n}\sum_{{\bf m}\in Z^2} f^n({\bf m})$. 
The shape of this distribution depends on the extent of the weighting
function $f$, 
hence on the correlation length, narrowing with longer correlation. 
This means we need a larger $W$ for stronger correlation in order to get 
comparable effects of disorder in the Hamiltonian.

Choosing $f({\bf m})=\Theta(R-|{\bf m}|)\,\exp(-|{\bf m}|^2/\eta^2)$ with
sufficiently large $R$, we obtain correlated random numbers with an 
approximately (due to the underlying lattice) Gaussian correlation
function
\begin{equation}
  K_{f}^{}({\bf m},{\bf m}')
  \,=\,
  \gamma^{\,2\big(\big[\frac{d_{x}}{2}\big]_{1}^{}
          +\big[\frac{d_{y}}{2}\big]_{1}^{}\big) }
  \
  \exp\Big(\!-\frac{|{\bf d}|^2}{2\eta^2} \Big)
  \,,
\end{equation}
with ${\bf d}={\bf m}-{\bf m}'$ and $[s]_{1}^{}\in [0,1)$ denoting 
the fractional part of the real number $s$. The $\gamma(\eta)$ are
0.0077, 0.70, and 0.97 for $\eta=0.3$, 0.7, and 1.0, respectively,
approaching $\gamma=1$ for $\eta\gtrsim 2$.
For the $g_{2}[f]$ we get 1.0, 0.42, 0.17, and 0.04 for 
$\eta=0.3$, 0.7, 1.0, and 2.0, respectively.

\begin{figure}[t]
 \begin{center}
  \includegraphics[width=7.5cm]{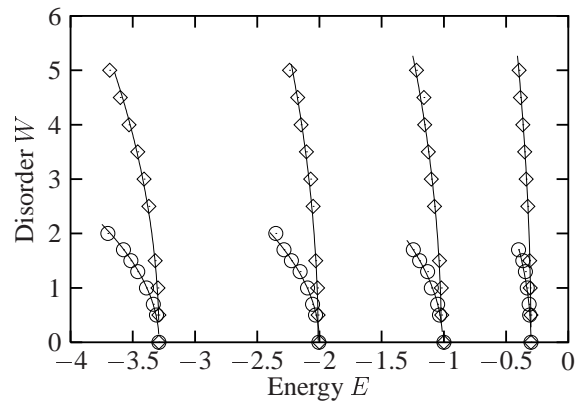}
  \end{center}
  \caption{The peak positions of the density of states $\rho(E,W,\eta)$ 
   in the disorder ($W$) energy ($E$) plane. The DOS peaks move outwards 
   (widening of the total TB band) with increasing
   disorder for a Gaussian correlation length $\eta=0.3$ 
   ({\large$\circ$}) and  $\eta=1.0$ ($\Diamond$).
   The solid lines are quadratic fits $\delta E_{n}\propto -W^2$.
 }
 \label{F_DOS_1}
\end{figure}

A Lorentzian like correlated random potential can be constructed by
choosing $f({\bf m})=\Theta(R-|{\bf m}|)\ \eta^2/(\eta^2+|{\bf
m}|^2)$. The corresponding correlation function is 
\[
  K_{f}({\bf m},{\bf m}')
  \,\approx\,
  \frac{1}{\sqrt{\big(\frac{|{\bf m}-{\bf m}'|}{2\eta}\big)^2+1}}
  \,+\, o(|{\bf m}-{\bf m}'|)
  \,.
\]
In this case we have $g_{2}[f]=2.9\cdot10^{-5}$ for $\eta=1.0$.

\section{The behavior of the density of states}
Independent of the disorder model a non-zero disorder potential has 
two main effects on the density of states (DOS). First, the narrow 
Harper bands that replace the totally degenerated Landau levels
of the continuum model on the lattice broaden nearly symmetrically.
The total width of a sub-band consists of the intrinsic Harper
broadening and the additional contribution due to the disorder potentials.
For large correlation length the disorder induced sub-band broadening 
decreases with increasing Landau level index until the finite
intrinsic width of the central Harper bands dominates. 
For small correlation length the disorder broadening
is nearly independent of the Landau level index. The remaining difference 
in the observed sub-band broadening is associated with the varying intrinsic
width of the unperturbed Harper bands. 
This usually leads to a slightly increasing width of the 
disorder broadened Harper bands with increasing proximity to the 
tight-binding (TB) band center.
The shape of these sub-bands depends on the correlation length, becoming
Gaussian for large $\eta$. Second, with increasing disorder strength
the center position of the sub-bands move in energy towards the edges 
of the TB band. This effect is not small and stronger for the outer sub-bands.

\begin{figure}
 \begin{center}
 \includegraphics[width=7.5cm]{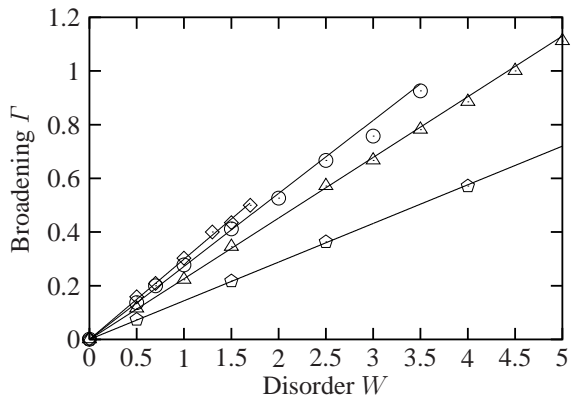}%
 \end{center}
 \caption[]{%
   The width $\mathit{\Gamma}$ of the disorder broadened   
   lowest Harper band vs. disorder strength $W$ 
   for different Gaussian correlation lengths
   $\eta=0.3$ ($\Diamond$), $0.7$ 
   (\raisebox{-.3ex}{\Large$\circ$}), 
   $1.0$ ({\scriptsize$\triangle$}), and $2.0$ (\pentagon).
   The solid lines are linear fits $\mathit{\Gamma}\propto W$. 
 }
 \label{F_DOS_2}
\end{figure}

The traces of the center positions of the disorder broadened Harper bands 
in the disorder-energy plane are plotted in Fig.~\ref{F_DOS_1}.
Because of the symmetry of the TB band, only the lower half is shown.  
The  widening of the total TB band is proportional to the 
square of the disorder strength within the resolvable range. 
The solid lines in the picture are fits $W\propto\sqrt{-\delta E_{n}}$ 
for each sub-band $n$, where $\delta E_{n}$ is the shift in energy 
of the respective DOS peak position due to the disorder $W$. We note that
$\delta E_{n}$ gets gradually smaller with increasing Landau band index $n$.
In contrast, the disorder broadening of each individual Harper band 
is only linear in the disorder strength, $\mathit{\Gamma}\propto W$.
This holds with good accuracy within the accessible range of disorders, 
independent of the correlation length as shown in Fig.~\ref{F_DOS_2}. 
The broadening of the Harper bands is stronger for shorter correlation length
of the disorder potential, however, taking the narrowing of the random
potential distribution function with increasing correlation into
account, we find stronger broadening for larger correlation length 
if the second moments of the random potential distribution functions
are set equal. 

 \begin{figure}[b]
 \begin{center}
 \includegraphics[width=7.5cm]{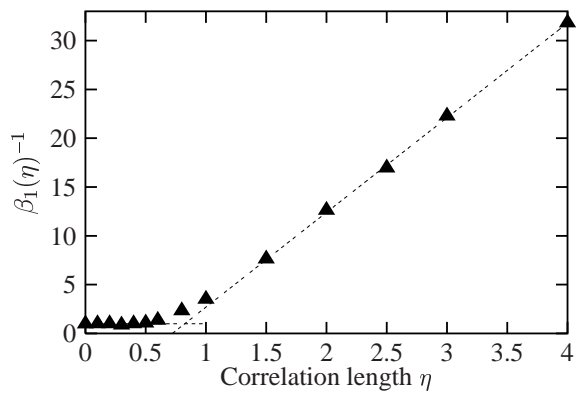}%
 \end{center}
 \caption{
   The inverse of the pre-factor $\beta_{1}$ of the quadratic shift,
   $\delta E_{1}=-\beta_{1}(\eta)\mathit{\Gamma}_{1}(W)^2$,  of
   the DOS peak position of the first sub-band 
   is shown as a function of the correlation length
   for Gaussian correlated disorder and fixed magnetic field $B=1/8$.
   The dashed lines are partial linear fits.
 }
 \label{F_BETA}
\end{figure}

The independence of the observed linearity from the correlation length 
enables us to define an {\em effective} disorder strength 
$W_{\mathrm{eff}}(W,\eta)$ which causes the same bandwidth
$\mathit{\Gamma}(W,\eta)=\mathit{\Gamma}(W_{\mathrm{eff}})$ 
of the lowest Harper band independent of the disorder model.
This measure for the disorder strength allows the direct comparison between
different correlations length and is also readily accessible from experimental 
data. For a given $W_{\mathrm{eff}}$ the spreading of the TB band decreases
with increasing correlation length.
For large correlation length ($\eta\gg 4$) the density of states  
can be described by the local distribution of the random potential folded 
with the unperturbed Harper band structure.
For Lorentzian and exponential correlation, not shown here, we 
find qualitatively the the same behavior of the DOS.
Our results are also in agreement with investigations on continuum models. 
\cite{AFS82,And84}

\begin{figure}[t]
 \begin{center}
  \includegraphics[width=7.5cm]{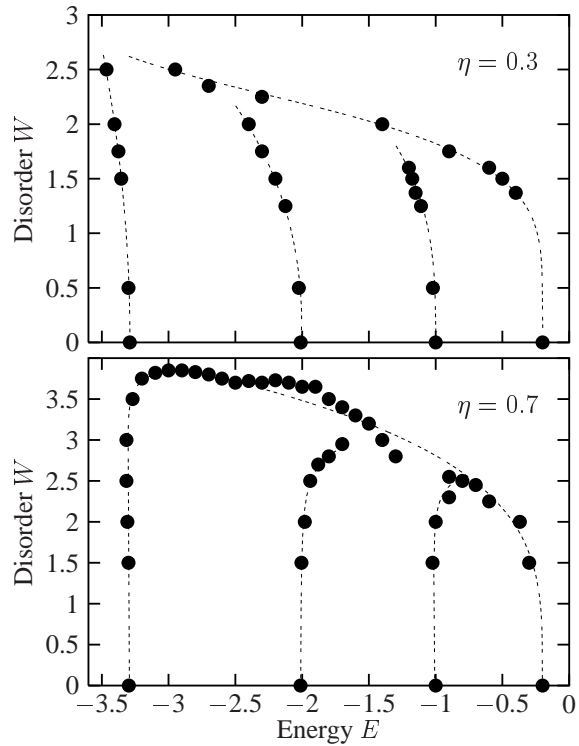}
 \end{center}
 \caption{The motion of the extended states with increasing 
   disorder in the disorder--energy plane is shown for the lowest four
   Harper bands. The states starting at $E=-3.3$, $E=-2$, $E=-1$ are
  the Chern states, while the anti-Chern states start at
  $E\approx-0.2$. The Gaussian correlation
   lengths are $\eta=0.3$ and $0.7$, respectively.
   The dashed lines are guides to the eye.  }
 \label{F_ES_1}
\end{figure}

Due to the linear broadening $\mathit{\Gamma}_{1}\propto W$ of 
the lowest sub-band and the quadratic shift
$\delta E_{1}\propto -W^2$ of the peak position of its DOS as a 
function of disorder strength, the energetical shift is consequently 
quadratic in the sub-band's width, 
$\delta E_{1} = -\beta_{1}(\eta)\ \mathit{\Gamma}_{1}(W)^2$.
As shown in Fig.~\ref{F_BETA}, the pre-factor $\beta_{1}(\eta)$ behaves
for large correlation length like $\beta_{1}\sim 1/\eta$, but saturates 
at a finite value for small correlation length.

\section{The behavior of the critical states}
The consideration of finite spatial correlations within the random
disorder potentials resolves  
the discrepancy between the expectations of levitation of 
current carrying states within the continuum model \cite{Khm84,Lau84} 
and the downward moving anti-Chern states observed in lattice models with 
white noise disorder. \cite{LXN96,XLSN96,SW97,SW98} We will show that 
both scenarios are only two special cases of the lattice model with 
correlated disorder that are associated with different spatial
correlation lengths of the disorder potential.

\begin{figure}
 \begin{center}
  \includegraphics[width=7.5cm]{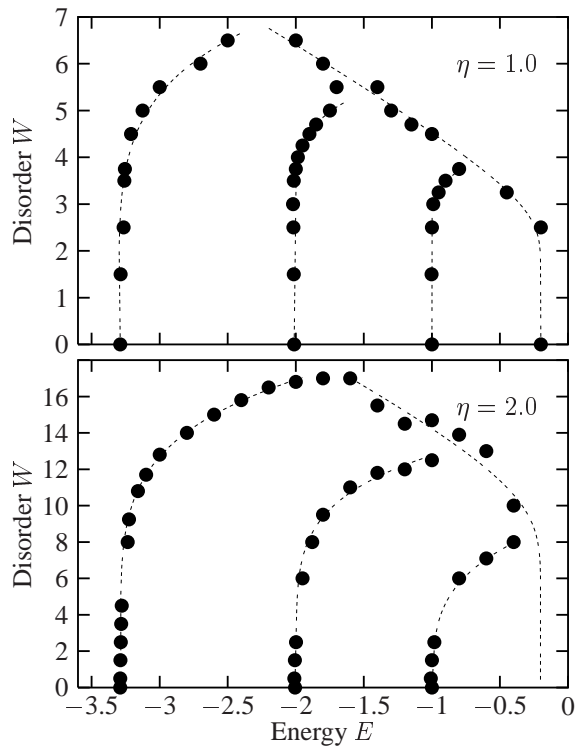}
 \end{center}
 \caption{The 
   same as in Fig.\ref{F_ES_1} with the exception that the  
   correlation lengths of the Gaussian correlated disorder potentials
   are $1.0$ and $2.0$, respectively.
}
\label{F_ES_1b}
\end{figure}

In Figs.~\ref{F_ES_1} and \ref{F_ES_1b} the traces of the critical 
states for Gaussian 
correlated disorder are shown for several correlation lengths $\eta$ 
in the disorder--energy plane. Again, the upper half of the 
tight-binding band is omitted as it is equal to the lower half 
mirrored at the line $E=0$.
For short correlation length $\eta=0.3$, 
we find no levitation of the lower sub-bands which, in fact, 
move absolutely down in energy following the center position of
the DOS peaks of the spreading TB band, though to a lesser extend.
The anti-Chern states which reside near the band center for the Harper model
without a disorder potential, rapidly move down in energy with increasing 
disorder strength, 
annihilating the Chern states from higher to lower Landau level index
until eventually the last extended state vanishes near the band edge.
For a slightly larger correlation length $\eta=0.7$, the lower sub-bands 
stop moving downwards in energy and remain at their unperturbed energies 
up to a quite large disorder strength. The anti-Chern states still move 
quickly down with increasing $W$, but less rapidly than for smaller $\eta$. 
Only shortly before getting annihilated by the anti-Chern the extended 
states from the lower sub-bands start to move to higher energy. Now the 
last extended state is destroyed slightly above the energy of the 
lowest unperturbed Harper band.

\begin{figure}[t]
 \begin{center}
  \includegraphics[width=7.75cm]{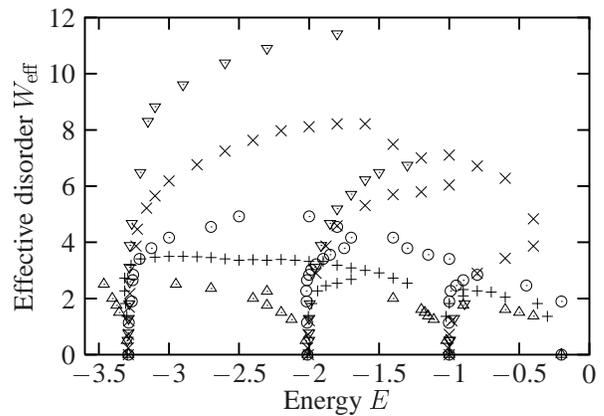}
 \end{center}
 \caption{The traces of the extended states (Chern and anti-Chern) vs. 
   energy for different Gaussian correlation length, 
   $\eta=0.3$ ({\scriptsize$\triangle$}), 
   $0.7$ (+), 
   $1.0$ ({\large$\circ$}), 
   $2.0$ ($\times$), and $4.0$ ($\triangledown$) 
   are compared using the effective disorder $W_{\mathrm{eff}}$ 
   defined by an equal broadening of the lowest Harper band.}
 \label{F_WoE_eff}
\end{figure}

For longer correlation length (see Fig.~\ref{F_ES_1b}) 
the extended states from the lower
sub-bands strongly move up in energy while the anti-Chern states 
still move towards the band edges. The energy where the last extended state
is destroyed when the lowest Chern and the anti-Chern states annihilate 
moves further towards the band center, passing the unperturbed 2nd
Harper band energy for $\eta\approx 2$. 
If the correlation length is increased further the anti-Chern states 
remain too close to the band center to be resolved anymore,  
while the critical states of the lower sub-bands float up in energy
clearly across the Landau gaps.
Thus, for large correlation length, i.e., for smooth disorder potentials, 
we reach the levitation picture anticipated from the continuum model.
From our data we see no indication for a merging of the critical states 
from the lower sub-bands in contrast to another report. \cite{SWW01}

For sufficiently large correlation lengths and disorder strengths that
produce equal widths of the lowest sub-band, the levitation of the 1st 
and 2nd critical states becomes slower with increasing $\eta$. 
This behavior is shown in Fig.~\ref{F_WoE_eff}.
Also, the levitation gets stronger with increasing Landau level index.
In contrast, the behavior is different for very small $\eta$ and is 
probably dominated by lattice effects. Similar results have been obtained 
for other disorder models with exponentially \cite{KPS01} and
Lorentzian like decaying correlation function.

\begin{figure}
 \begin{center}
  \includegraphics[width=7.5cm]{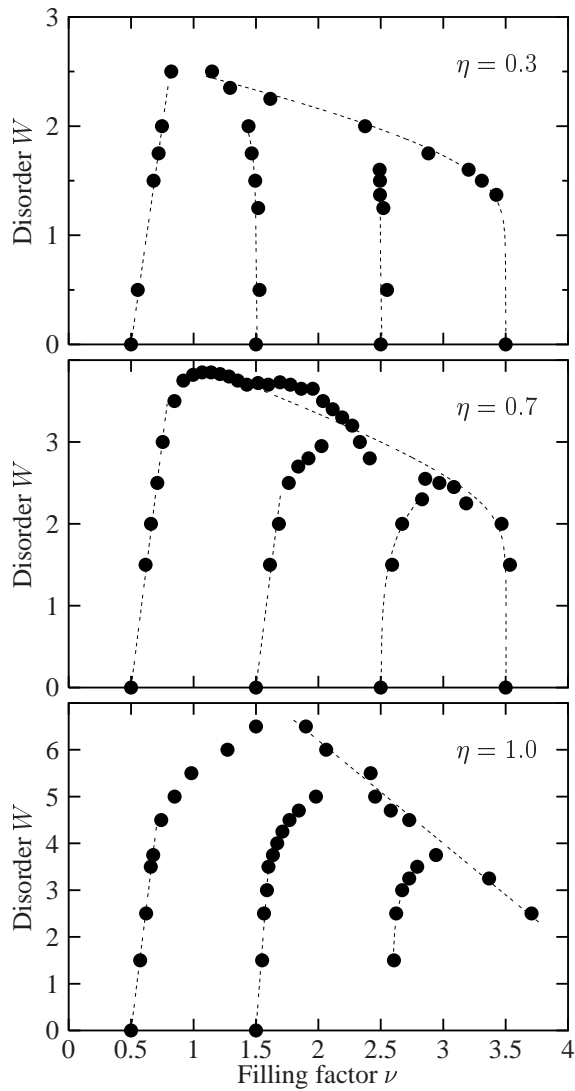}
 \end{center}
 \caption{The three figures show the behavior of the extended states in
   the disorder -- filling factor plane for Gaussian correlation with
   $\eta=0.3$, $0.7$, and $1.0$. The dashed lines are guides to the eye. }
 \label{F_ES_2}
\end{figure}

In Fig.~\ref{F_ES_2} the traces of the extended states' positions with 
respect to the filling factor are plotted. They clearly differ from the 
energy dependence (Figs.~\ref{F_ES_1} and \ref{F_ES_1b}) since they contain 
the behavior of the 
extended states and the total DOS combined. As a function of the filling 
factor the observed levitation is even stronger because of the spreading
of the tight-binding band and the additional rather small contribution 
originating from the accumulation of the density of states   
at a particular band that comes from sub-bands above
(density floating \cite{LXN96,XLSN96,PBS98}). 
With respect to $\nu$, the extended states from the lower band always 
levitate to higher fillings, \cite{PS02a} even in the uncorrelated case.
The prominent linear slope of the lower extended states' traces for not 
too strong disorder is essentially an effect of the DOS behavior. 
The widening of the TB band $\propto W^2$ and the broadening of the Harper
sub-bands $\propto W$ together with either the shift down in energy $\propto
W^2$ ($\eta\le 0.3$) or the absence of any movement ($\eta\ge 0.7$) 
leads for a qualitatively Gaussian like shape of the sub-bands to a 
prominent linear term $\propto W$ in the filling factor dependence for
small $W$. This feature gets less pronounced for larger $\eta$.

\begin{figure}
 \begin{center}
  \includegraphics[width=7.5cm]{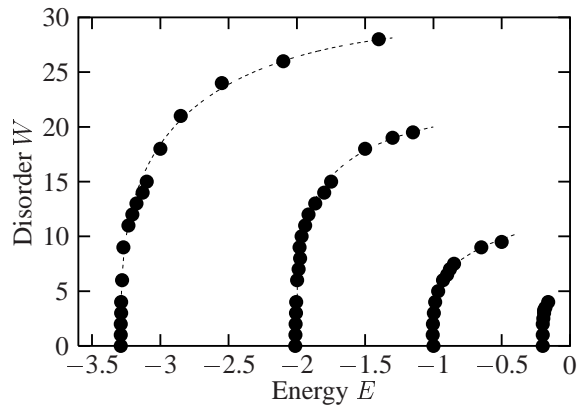}
 \end{center}
 \caption{The behavior of the current carrying states for a long-range 
   Lorentzian like 
   correlated background potential with correlation length $\eta=1.0$ 
   is shown in the disorder--energy plane.
   The downward movement of the anti-Chern cannot be resolved anymore.
   The dashed lines are guides to the eye. }
 \label{F_ES_3E}
\end{figure}

\begin{figure}[b]
 \begin{center}
  \includegraphics[width=7.5cm]{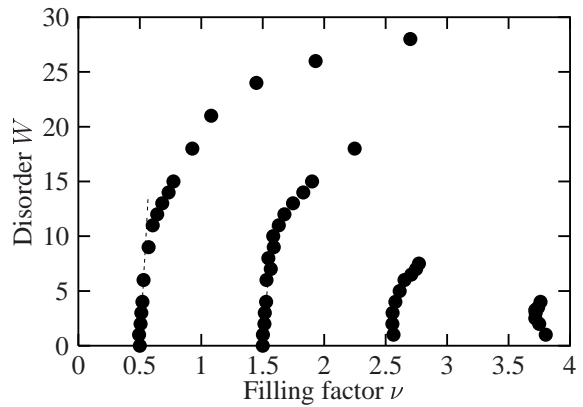}
 \end{center}
 \caption{The behavior of the current carrying states for a long-range 
   Lorentzian like 
   correlated background potential with correlation length $\eta=1.0$ 
   is shown in the disorder--filling factor plane.
   The dashed lines indicate the linear slope at small $W$. }
 \label{F_ES_3N}
\end{figure}

For the disorder model with long-range Lorentzian like correlations, 
we find qualitatively the same behavior as for the Gaussian correlated 
random potential. The results are also compatible with those obtained
previously for exponentially correlated disorder. \cite{KPS01}
In Figs.~\ref{F_ES_3E} and \ref{F_ES_3N} the traces of the extended 
states are shown vs. energy $E$ and filling factor $\nu$, respectively, 
for a correlation length $\eta=1.0$. The levitation of the
extended states from the lowest band across one and a half Landau gap
in energy and even across two Landau gaps w.r.t. filling factor is 
clearly seen.
Also, the second band's extended states can be traced nearly across one 
Landau gap in this case. The anti-Chern states from the TB band center
are moving down too slowly with increasing disorder strength $W$ to be 
resolved in this case.

\section{Discussion of the results}
The physical problem investigated here is governed by two fundamental
scales, the energy scale $\hbar\omega_{c}/V$, which for the lattice model 
corresponds to $4\pi\alpha_{B}$, and the magnetic length 
$l_{B}/a=(2\pi\alpha_{B})^{1/2}$, where $\alpha_{B}=p/q$ is the number of 
flux quanta per lattice plaquette $a^2$. For strong magnetic flux
densities $B=\alpha_{B} (h/e) a^{-2}$, $l_{B}$ gets smaller 
than the lattice constant $a$, a domain not considered in the present
work. Here we have chosen $\alpha_{B}=1/8$ which guarantees, in connection 
with spatially correlated disorder potentials and our presently available 
computer capacity, the ability to numerically resolve the entire ranges 
of energy and disorder values as shown above.

To discuss the influence of the correlation length of the disorder
potentials one needs a practicable measure that allows the comparing 
of the various results. One choice is the second moment of the
disorder potentials which is independent of $B$ and can be viewed
essentially as fixed for a given physical sample. 
A second possibility is the $B$-dependent disorder broadening of 
the sub-bands, $\mathit{\Gamma} \propto W$, which would be more advantageous 
in comparing with experiments. This method has been used in 
Fig.~\ref{F_WoE_eff} where traces of the critical states
have been plotted versus an effective disorder strength. 
However, for weak magnetic field and/or
strong disorder this quantity may not easily be accessible.  

In considering the observed energetical movement of the current carrying 
states $E_{\mathrm{c}}$, one has to distinguish two contributions. One is 
related to the outward shift of the energetic position of the individual 
sub-bands of the DOS with increasing disorder strength, $\delta E_{n} \propto
W^2$, which leads to the linear dependence on the filling factor $\nu$. 
The second contribution is the absolute energetical shift, 
$\delta E_{\mathrm{c}}=E_{\mathrm{c}}-E_{\mathrm{c}}(W=0)$, 
of the critical states (Chern and anti-Chern states) across the Landau gaps.
Both contributions depend on the correlation length of the disorder
potentials and on the strength of the magnetic field. To compare with 
the energy shift proposed within the levitation scenario in the continuum 
limit, \cite{Khm84,Lau84} $\delta E \propto (n+1/2) (\omega_{c}\tau)^{-2}$, 
the absolute energetical shift $\delta E_{\mathrm{c}}$ has to be 
considered. We find a qualitative agreement with this relation for larger 
correlation lengths $\eta$. For comparison with experiments, the calculated 
filling factor dependence of $\delta E_{\mathrm{c}}$ (see Figs.~\ref{F_ES_2} 
and \ref{F_ES_3N}) seems to be better suited.

The levitation of the energy of the lowest current carrying states with
respect to the sub-band peak of the density of states has recently been 
proposed to follow the relation \cite{PS02b}
\begin{equation}
\frac{\delta E_{\mathrm{eff}}}{\hbar\omega_{c}}\sim \frac{l_{B}}{\eta}
\Big(\frac{\mathit{\Gamma}}{\hbar\omega_{c}}\Big)^2.
\label{deff}
\end{equation}
The definition of an effective shift $\delta E_{\mathrm{eff}}$, given by 
the energy difference between position of the DOS peaks and the respective 
current carrying states \cite{YB96,PS02a} is only practical as long 
as the DOS peaks can clearly be resolved which, however, gets 
increasingly problematic for small $B$ and strong $W$. Therefore, 
it is clear, that this definition can, as a matter of principle, 
not be applied for the strong levitation 
across the Landau gaps seen in our investigation. In contrast, the
resolution of the energetical position of the critical states can
continously be enhanced by applying larger system sizes.
Our results also show that even in the weak levitation regime where
the energetical shift of the critical states $\delta E_{\mathrm{eff}}$ 
is essentially due to the shift of the DOS peak $\delta E_{1}$, the 
proposed linear relation $(\delta E_{\mathrm{eff}})^{-1}\sim \eta$ 
only holds over a limited range of $\eta$. It strongly deviates for 
small $\eta$ (see Fig.~\ref{F_BETA}) where $\delta E_{1}=
-\beta_{1}(\eta) \mathit{\Gamma}^2$ saturates.

\begin{figure}
 \begin{center}
  \includegraphics[width=7.5cm]{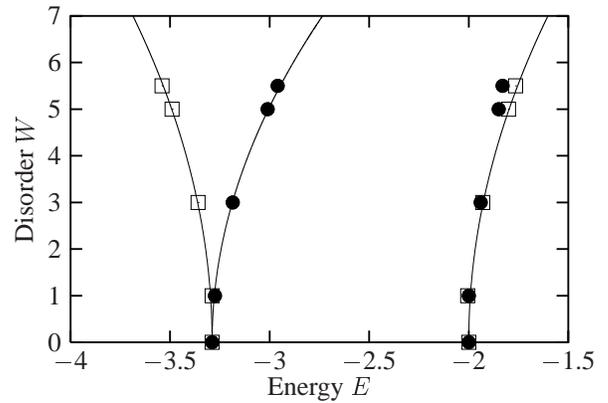}
 \end{center}
 \caption{The energetical positions of the DOS peaks ($\Box$) 
  and the positions of the critical states
  ({\large$\bullet$}) of the lowest two sub-bands for the projected 
   disordered Harper model with correlation
   length $\eta=1$ and magnetic field $B=1/8$
   are shown vs.\ energy $E$.
   The solid lines are the quadratic fits $\delta E\propto W^2$.}
\label{F_2band}
\end{figure}

\begin{figure}[b]
 \begin{center}
  \includegraphics[width=7.5cm]{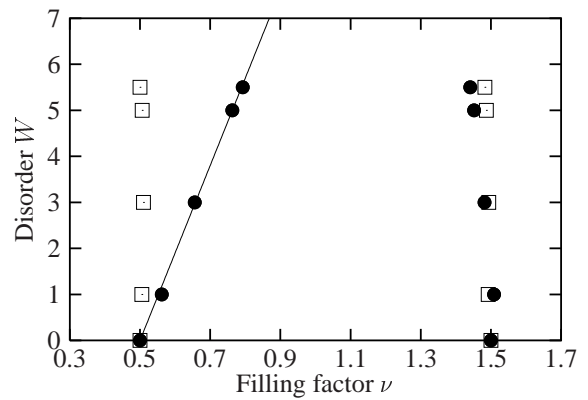}
 \end{center}
 \caption{For the projected disordered Harper model with correlation
  length $\eta=1$ and magnetic field $B=1/8$ 
  the positions of the DOS peaks ($\Box$) 
  and the positions of the critical states
  ({\large$\bullet$}) of the lowest two sub-bands 
   are shown vs.\ filling factor $\nu$. 
   The solid line is a linear fit $\delta\nu\propto W$.  }
\label{F_2band_nu}
\end{figure}

In the large disorder regime, where the strong levitation is dominated
by the absolute energetical shift, even the relation 
$\delta E_{\mathrm{c}} \sim W^2$ no longer holds. Instead, we clearly
observe a dependence with an exponent larger than 2. This behavior
may change with decreasing $B$, but, due to the limited system size,  
it was not possible to resolve this problem yet.
Finally, we did not see any merging of the levitating states, therefore
true direct transitions are not possible. However, for smaller
magnetic fields the critical states may get very close so that their 
resolution is not possible any more.

\section{Projected disordered Harper model}
In order to get an idea of a possible mechanism that leads to the 
levitation of the critical states, we furthermore investigated a 
disordered Harper model projected onto the lowest few Landau levels. 
This was done by a 
projection of the highly non-diagonal spatial disorder potential 
within the base spanned by the unperturbed Harper eigenstates.
Starting by using only the two lowest Harper bands, we could 
reproduce the widening
of the total band as well as the levitation of the critical states of the 
lowest sub-band which is shown in Fig.~\ref{F_2band} for a Gaussian
correlation length $\eta=1$. 
Since there are only two levels repelling each other, the center of 
the 2nd DOS peak moves to higher energies, in contrast to the 
full model. This widening leads to a shift of both sub-bands $\propto W^2$ 
as in the full model.
The levitation of the lowest critical state is clearly visible while 
the extended states of the 2nd level are essentially following the DOS peak.
Different from the full model, the levitation of the 1st level is
with good accuracy $\propto W^2$ over the entire resolvable range. 
The levitation w.r.t. filling factor is again linear in $W$
(Fig.~\ref{F_2band_nu}). 
Further, we do not see any floating down of extended states for smaller 
correlation length as it has been observed for instance for 
Gaussian correlation with $\eta=0.3$ in the full model (see Fig.~\ref{F_ES_1}).

\begin{figure}[t]
 \begin{center}
  \includegraphics[width=7.5cm]{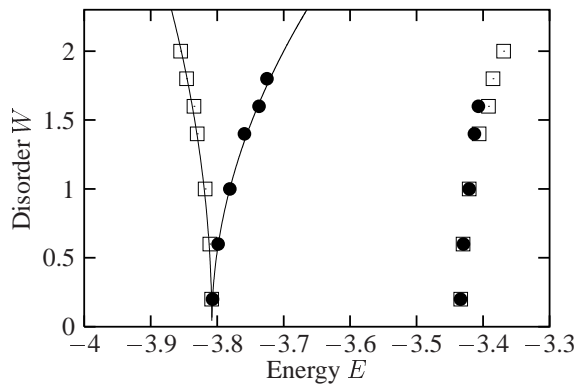}
 \end{center}
 \caption[]{For the projected disordered Harper model with correlation
  length $\eta=1$ and magnetic field $B=1/32$ 
  the positions of the DOS peaks ($\Box$) 
  and the positions of the critical states
  ({\large$\bullet$}) of the lowest two sub-bands 
   are shown vs.\ energy $E$. 
   The solid lines are quadratic fits $\delta E\propto W^2$. }
\label{F_2band_32}
\end{figure}

Dropping the matrix elements describing the inter-band interaction  of
the disorder part in the Hamiltonian containing only two sub-bands,
completely destroys the levitation as well as the widening of the
total tight-binding band, while preserving the localization properties.

For a smaller magnetic field $B=1/32$ the behavior of the disordered
Harper model projected onto the lowest two Harper bands shown in
Fig.~\ref{F_2band_32} is qualitatively the same. However, the ratio of
upward movement of the first sub-band's extended states and the
downward movement of the corresponding DOS peak position  changes with
magnetic field. The shape of the traces remains purely quadratic in
any case.

\begin{figure}
  \includegraphics[width=7.5cm]{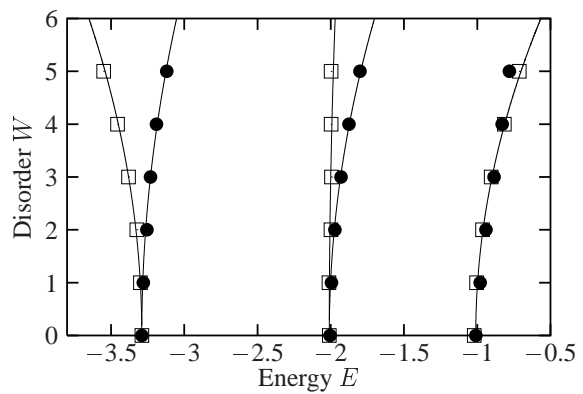}
 \caption[]{For the projected disordered Harper model with correlation
  length $\eta=1$ and magnetic field $B=1/8$ 
  the positions of the DOS peaks ($\Box$) 
  and the positions of the critical states
  ({\large$\bullet$}) of the lowest three sub-bands 
   are shown vs.\ energy $E$. 
   The solid lines are the quadratic fits $\delta E\propto W^2$. }
\label{F_3band}
\end{figure}

The addition of the third unperturbed Harper band to the projected
model adds a qualitatively new behavior which is shown in
Fig.~\ref{F_3band}.  The DOS peak position of the second band remains
nearly at the same energy, those of the first and third bands are
moving outwards,  while the extended states from the first {\em and
second} bands move to higher  energies and filling factors with
increasing disorder strength.  In the third band, the energy of the
extended states essentially follows the DOS peak position. There is
no indication for any absolute levitation of these extended states.  
This corresponds to the behavior of the second band in the two-band 
model shown above. The levitation of the lowest band's extended states 
is slowed down by the addition of a third band compared to the 
two-band model. Now it resembles closer the trace known from the full
disordered Harper model (Fig.~\ref{F_ES_1}).  We conclude, that in
order to see levitation of a particular Harper band's  extended states
the existence of at least one adjacent band at higher energy is
required.

\section{Conclusions}
The influence of spatially correlated disorder potentials on the
levitation of current carrying states has been investigated for a
quantum Hall system described by a two-dimensional lattice model.
We considered what happens when the strength $W$ of the disorder 
potentials is increased so that neighboring Landau bands start to overlap. 
Here, we showed our findings obtained mainly for spatially correlated 
disorder potentials which were generated with Gaussian correlations 
characterized by a correlation length $\eta$, but results for a 
Lorentzian like correlated disorder have also been presented. 
They both qualitatively agree with the previously obtained outcome 
for exponentially correlated disorder. \cite{PS01,KPS01}
Thus, the results obtained previously are not the consequence
of the special exponentially correlated disorder model, but
fit into a picture of a more general disordered lattice model which 
for long-range correlated disorder potentials agrees qualitatively 
with what is expected from the continuum model without a periodic 
background potential.

We numerically calculated the density of states $\rho(E,W,\eta)$ and studied 
the scaling of the localization length $\lambda_{M}(E,W,\eta)$ for various 
energies $E$, disorder strengths $W$, and system widths $M$ 
from which the position of the critical states were extracted. 
Depending on the correlation length, we found three different cases.
For small correlation length $\eta\lesssim 0.3$  no levitation 
of the Chern states, but their annihilation by downward moving 
anti-Chern states has been observed.
For $\eta\gtrsim 2.0$ the Chern states
float up in energy, at least across one Landau gap. 
No effect of the anti-Chern states could be detected anymore 
for $\eta \ge 4$ where the anti-Chern remain close to the TB band
center. In the latter case a levitation scenario as expected from 
the continuum model without a periodic potential has
been observed in our lattice model. 
In the intermediate case
$0.7\lesssim\eta\lesssim 1.0$ both the levitation of the Chern as well
as the outwards moving anti-Chern states can be noticed.
As a function of the filling factor, the shift of the lower critical 
states is always positive, independent of the correlation length.
One has to bear in mind that the levitation is always accompanied by
the spreading of the tight-binding (TB) band, which is not negligible in
particular for small $\eta$. We also have to mention that in contrast
to a recent report~\cite{SWW01} no merging of current carrying states was
observed in our investigations. Therefore, no genuine direct transitions
are possible within the present model. Of course, also the suggested 
annihilation mechanism of the current carrying states by anti-Chern 
states has to be discarded because this scenario is a special feature 
of the lattice model that cannot serve as an explanation for the 
experimental observations of a seemingly direct transition.

Besides these results that confirm and complement our previous findings
for a model with exponentially distributed spatially correlated disorder 
potentials, we obtained the following new conclusions. 
\textit{i}) The introduction of an effective disorder $W_{\mathrm{eff}}$, 
which is defined by an equal broadening $\mathit{\Gamma}$ of the sub-bands, 
allows a general discussion of the influence on the correlation length $\eta$. 
This is advantageous when comparing with other disorder models and experiment. 
The definition was possible only because of the observed linear broadening 
$\mathit{\Gamma}_1 \propto W$ for different Gaussian correlation 
lengths $\eta$.
\textit{ii}) The observed movement in energy of the current carrying states 
$E_{\mathrm{c}}$ consists of at least two contributions, the shift 
of the density of states peaks $\delta E_{n}\propto -W^2$, and the absolute 
energetical shift which for low disorders follows a relation 
$\delta E_{\mathrm{c}} \propto W^2$. 
In the strong levitation regime at larger $W$, the absolute shift 
$\delta E_{\mathrm{c}}$ and therefore also the observed $E_{\mathrm{c}}$ 
move faster than $ W^2$. The peak of the lowest sub-band follows 
$\delta E_{1}\propto-\beta_{1}(\eta) W^2$ with $\beta_1(\eta)\sim 1/\beta$ 
for $\eta$ not too small, but saturates in the limit $\eta\to 0$.
\textit{iii}) The definition of an effective energetical shift of the current 
carrying states, i.e., $\delta E_{\mathrm{eff}} \sim |E_{1}-E_{\mathrm{c}}|$ 
for the lowest sub-band, is practicable only for small disorder strength 
where the DOS peaks can be resolved. Therefore, it does neither work in 
the medium and nor in the strong levitation regime across the Landau gap 
as observed in our work. Also, the proposed relation (Eq.~\ref{deff})
does not hold for small $\eta$ where $(\beta(\eta))^{-1}$ is no longer  
linear in $\eta$.
\textit{iv})
Our study of the truncated disordered Harper model, which did not
include the anti-Chern states, showed the levitation still to exist,
provided the reduced model contains at least the coupling between 
the sub-band considered and the adjacent sub-band at higher energy. 
The spreading of the TB band as well as the levitation of 
the lowest critical states could already be detected in a model 
projected only onto the two lowest unperturbed Harper bands.
Dropping the inter-sub-band interaction part of the correlated disorder
potentials destroyed the levitation and the spreading of the TB band
while preserving the localization properties. 
These results clearly show that the microscopic origin of the critical
states' floating up in energy originates in the matrix elements of the
disorder potentials constructed from eigenstates of consecutive 
Landau levels. In the case of the truncated lattice model the position
of the critical states was extracted using the method of level statistics.

To summarize,
we presented a comprehensive picture of the levitation of the current 
carrying states in the presence of spatially correlated disorder potentials.
The behavior cannot be described by a simple general relation of the 
energetical shift as a function of magnetic field and disorder strength. 
However, empirical relations have been found 
which hold in certain ranges of the disorder and correlation length
parameters. Further investigations, particularly at lower magnetic flux 
densities, are needed to clarify the complete functional dependences. 
Our results for the truncated disordered Haper model provide first
clues for a microscopical understanding of the levitation of the 
current carrying states at the quantum Hall to insulator transition.

\begin{acknowledgments}
We acknowledge stimulating discussions with A. L. C. Pereira and P. A. Schulz. 
This work was supported in part by the Schwerpunktpro\-gramm 
``Quantum-Hall-Systeme'' of the German Science Foundation (DFG).
\end{acknowledgments}

\bibliographystyle{apsrev}

\begin{thebibliography}{43}
\expandafter\ifx\csname natexlab\endcsname\relax\def\natexlab#1{#1}\fi
\expandafter\ifx\csname bibnamefont\endcsname\relax
  \def\bibnamefont#1{#1}\fi
\expandafter\ifx\csname bibfnamefont\endcsname\relax
  \def\bibfnamefont#1{#1}\fi
\expandafter\ifx\csname citenamefont\endcsname\relax
  \def\citenamefont#1{#1}\fi
\expandafter\ifx\csname url\endcsname\relax
  \def\url#1{\texttt{#1}}\fi
\expandafter\ifx\csname urlprefix\endcsname\relax\def\urlprefix{URL }\fi
\providecommand{\bibinfo}[2]{#2}
\providecommand{\eprint}[2][]{\url{#2}}

\bibitem[{\citenamefont{Jiang et~al.}(1993)\citenamefont{Jiang, Johnson, Wang,
  and Hannahs}}]{JJWH93}
\bibinfo{author}{\bibfnamefont{H.~W.} \bibnamefont{Jiang}},
  \bibinfo{author}{\bibfnamefont{C.~E.} \bibnamefont{Johnson}},
  \bibinfo{author}{\bibfnamefont{K.~L.} \bibnamefont{Wang}}, \bibnamefont{and}
  \bibinfo{author}{\bibfnamefont{S.~T.} \bibnamefont{Hannahs}},
  \bibinfo{journal}{Phys. Rev. Lett.} \textbf{\bibinfo{volume}{71}},
  \bibinfo{pages}{1439} (\bibinfo{year}{1993}).

\bibitem[{\citenamefont{Shashkin et~al.}(1993)\citenamefont{Shashkin,
  Kravchenko, and Dolgopolov}}]{SKD93}
\bibinfo{author}{\bibfnamefont{A.~A.} \bibnamefont{Shashkin}},
  \bibinfo{author}{\bibfnamefont{G.~V.} \bibnamefont{Kravchenko}},
  \bibnamefont{and} \bibinfo{author}{\bibfnamefont{V.~T.}
  \bibnamefont{Dolgopolov}}, \bibinfo{journal}{JETP Lett.}
  \textbf{\bibinfo{volume}{58}}, \bibinfo{pages}{220} (\bibinfo{year}{1993}).

\bibitem[{\citenamefont{Wang et~al.}(1994)\citenamefont{Wang, Clark, Spencer,
  Mack, and Kirk}}]{Wea94}
\bibinfo{author}{\bibfnamefont{T.} \bibnamefont{Wang}},
  \bibinfo{author}{\bibfnamefont{K.~P.} \bibnamefont{Clark}},
  \bibinfo{author}{\bibfnamefont{G.~F.} \bibnamefont{Spencer}},
  \bibinfo{author}{\bibfnamefont{A.~M.} \bibnamefont{Mack}}, \bibnamefont{and}
  \bibinfo{author}{\bibfnamefont{W.~P.} \bibnamefont{Kirk}},
  \bibinfo{journal}{Phys. Rev. Lett.} \textbf{\bibinfo{volume}{72}},
  \bibinfo{pages}{709} (\bibinfo{year}{1994}).

\bibitem[{\citenamefont{Hughes et~al.}(1994)\citenamefont{Hughes, Nicholls,
  Frost, Linfield, Pepper, Ford, Ritchie, Jones, Kogan, and Kaveh}}]{Hea94a}
\bibinfo{author}{\bibfnamefont{R.~J.~F.} \bibnamefont{Hughes}},
  \bibinfo{author}{\bibfnamefont{J.~T.} \bibnamefont{Nicholls}},
  \bibinfo{author}{\bibfnamefont{J.~E.~F.} \bibnamefont{Frost}},
  \bibinfo{author}{\bibfnamefont{E.~H.} \bibnamefont{Linfield}},
  \bibinfo{author}{\bibfnamefont{M.}~\bibnamefont{Pepper}},
  \bibinfo{author}{\bibfnamefont{C.~J.~B.} \bibnamefont{Ford}},
  \bibinfo{author}{\bibfnamefont{D.~A.} \bibnamefont{Ritchie}},
  \bibinfo{author}{\bibfnamefont{G.~A.~C.} \bibnamefont{Jones}},
  \bibinfo{author}{\bibfnamefont{E.}~\bibnamefont{Kogan}}, \bibnamefont{and}
  \bibinfo{author}{\bibfnamefont{M.}~\bibnamefont{Kaveh}}, \bibinfo{journal}{J.
  Phys. Condens. Matter} \textbf{\bibinfo{volume}{6}}, \bibinfo{pages}{4763}
  (\bibinfo{year}{1994}).

\bibitem[{\citenamefont{Glozman et~al.}(1995)\citenamefont{Glozman, Johnson,
  and Jiang}}]{GJJ95}
\bibinfo{author}{\bibfnamefont{I.}~\bibnamefont{Glozman}},
  \bibinfo{author}{\bibfnamefont{C.~E.} \bibnamefont{Johnson}},
  \bibnamefont{and} \bibinfo{author}{\bibfnamefont{H.~W.} \bibnamefont{Jiang}},
  \bibinfo{journal}{Phys. Rev. Lett.} \textbf{\bibinfo{volume}{74}},
  \bibinfo{pages}{594} (\bibinfo{year}{1995}).

\bibitem[{\citenamefont{Khmelnitskii}(1984)}]{Khm84}
\bibinfo{author}{\bibfnamefont{D.~E.} \bibnamefont{Khmelnitskii}},
  \bibinfo{journal}{Phys. Lett.} \textbf{\bibinfo{volume}{106A}},
  \bibinfo{pages}{182} (\bibinfo{year}{1984}).

\bibitem[{\citenamefont{Laughlin}(1984)}]{Lau84}
\bibinfo{author}{\bibfnamefont{R.~B.} \bibnamefont{Laughlin}},
  \bibinfo{journal}{Phys. Rev. Lett} \textbf{\bibinfo{volume}{52}},
  \bibinfo{pages}{2304} (\bibinfo{year}{1984}).

\bibitem[{\citenamefont{Kivelson et~al.}(1992)\citenamefont{Kivelson, Lee, and
  Zhang}}]{KLZ92}
\bibinfo{author}{\bibfnamefont{S.}~\bibnamefont{Kivelson}},
  \bibinfo{author}{\bibfnamefont{D.-H.} \bibnamefont{Lee}}, \bibnamefont{and}
  \bibinfo{author}{\bibfnamefont{S.-C.} \bibnamefont{Zhang}},
  \bibinfo{journal}{Phys. Rev. B} \textbf{\bibinfo{volume}{46}},
  \bibinfo{pages}{2223} (\bibinfo{year}{1992}).

\bibitem[{\citenamefont{Shahar et~al.}(1995)\citenamefont{Shahar, Tsui, and
  Cunningham}}]{STC95}
\bibinfo{author}{\bibfnamefont{D.}~\bibnamefont{Shahar}},
  \bibinfo{author}{\bibfnamefont{D.~C.} \bibnamefont{Tsui}}, \bibnamefont{and}
  \bibinfo{author}{\bibfnamefont{J.~E.} \bibnamefont{Cunningham}},
  \bibinfo{journal}{Phys. Rev. B} \textbf{\bibinfo{volume}{52}},
  \bibinfo{pages}{R14372} (\bibinfo{year}{1995}).

\bibitem[{\citenamefont{Lee et~al.}(1998)\citenamefont{Lee, Chang, Suen, and
  Lin}}]{LCSL98}
\bibinfo{author}{\bibfnamefont{C.~H.} \bibnamefont{Lee}},
  \bibinfo{author}{\bibfnamefont{Y.~H.} \bibnamefont{Chang}},
  \bibinfo{author}{\bibfnamefont{Y.~W.} \bibnamefont{Suen}}, \bibnamefont{and}
  \bibinfo{author}{\bibfnamefont{H.~H.} \bibnamefont{Lin}},
  \bibinfo{journal}{Phys. Rev. B} \textbf{\bibinfo{volume}{58}},
  \bibinfo{pages}{10629} (\bibinfo{year}{1998}).

\bibitem[{\citenamefont{Hilke et~al.}(2000)\citenamefont{Hilke, Shahar, Song,
  Tsui, and Xie}}]{Hea00}
\bibinfo{author}{\bibfnamefont{M.}~\bibnamefont{Hilke}},
  \bibinfo{author}{\bibfnamefont{D.}~\bibnamefont{Shahar}},
  \bibinfo{author}{\bibfnamefont{S.~H.} \bibnamefont{Song}},
  \bibinfo{author}{\bibfnamefont{D.~C.} \bibnamefont{Tsui}}, \bibnamefont{and}
  \bibinfo{author}{\bibfnamefont{Y.~H.} \bibnamefont{Xie}},
  \bibinfo{journal}{Phys. Rev. B} \textbf{\bibinfo{volume}{62}},
  \bibinfo{pages}{6940} (\bibinfo{year}{2000}).

\bibitem[{\citenamefont{Dultz et~al.}(1998)\citenamefont{Dultz, Jiang, and
  Schaff}}]{DJS98}
\bibinfo{author}{\bibfnamefont{S.~C.} \bibnamefont{Dultz}},
  \bibinfo{author}{\bibfnamefont{H.~W.} \bibnamefont{Jiang}}, \bibnamefont{and}
  \bibinfo{author}{\bibfnamefont{W.~J.} \bibnamefont{Schaff}},
  \bibinfo{journal}{Phys. Rev. B} \textbf{\bibinfo{volume}{58}},
  \bibinfo{pages}{7532} (\bibinfo{year}{1998}).

\bibitem[{\citenamefont{{Hanein et al.}}(1999)}]{HaN99}
\bibinfo{author}{\bibfnamefont{Y.}~\bibnamefont{{Hanein et al.}}},
  \bibinfo{journal}{Nature} \textbf{\bibinfo{volume}{400}},
  \bibinfo{pages}{735} (\bibinfo{year}{1999}).

\bibitem[{\citenamefont{Yasin et~al.}(2002)\citenamefont{Yasin, Simmons,
  Lumpkin, Clark, Pfeifer, and West}}]{Yea02}
\bibinfo{author}{\bibfnamefont{C.~E.} \bibnamefont{Yasin}},
  \bibinfo{author}{\bibfnamefont{M.~Y.} \bibnamefont{Simmons}},
  \bibinfo{author}{\bibfnamefont{N.~E.} \bibnamefont{Lumpkin}},
  \bibinfo{author}{\bibfnamefont{R.~G.} \bibnamefont{Clark}},
  \bibinfo{author}{\bibfnamefont{L.~N.} \bibnamefont{Pfeifer}},
  \bibnamefont{and} \bibinfo{author}{\bibfnamefont{K.~W.} \bibnamefont{West}},
  \bibinfo{journal}{cond-mat/0204519} pp. \bibinfo{pages}{1--4}
  (\bibinfo{year}{2002}).

\bibitem[{\citenamefont{Shahbazyan and Raikh}(1995)}]{SR95}
\bibinfo{author}{\bibfnamefont{T.~V.} \bibnamefont{Shahbazyan}}
  \bibnamefont{and} \bibinfo{author}{\bibfnamefont{M.~E.} \bibnamefont{Raikh}},
  \bibinfo{journal}{Phys. Rev. Lett.} \textbf{\bibinfo{volume}{75}},
  \bibinfo{pages}{304} (\bibinfo{year}{1995}).

\bibitem[{\citenamefont{Gramada and Raikh}(1996)}]{GR96}
\bibinfo{author}{\bibfnamefont{A.}~\bibnamefont{Gramada}} \bibnamefont{and}
  \bibinfo{author}{\bibfnamefont{M.~E.} \bibnamefont{Raikh}},
  \bibinfo{journal}{Phys. Rev. B} \textbf{\bibinfo{volume}{54}},
  \bibinfo{pages}{1928} (\bibinfo{year}{1996}).

\bibitem[{\citenamefont{Haldane and Yang}(1997)}]{HY97}
\bibinfo{author}{\bibfnamefont{F.~D.~M.} \bibnamefont{Haldane}}
  \bibnamefont{and} \bibinfo{author}{\bibfnamefont{K.}~\bibnamefont{Yang}},
  \bibinfo{journal}{Phys. Rev. Lett.} \textbf{\bibinfo{volume}{78}},
  \bibinfo{pages}{298} (\bibinfo{year}{1997}).

\bibitem[{\citenamefont{Fogler}(1998)}]{Fog98}
\bibinfo{author}{\bibfnamefont{M.~M.} \bibnamefont{Fogler}},
  \bibinfo{journal}{Phys. Rev. B} \textbf{\bibinfo{volume}{57}},
  \bibinfo{pages}{11947} (\bibinfo{year}{1998}).

\bibitem[{\citenamefont{Liu et~al.}(1996)\citenamefont{Liu, Xie, and
  Niu}}]{LXN96}
\bibinfo{author}{\bibfnamefont{D.~Z.}~\bibnamefont{Liu}},
  \bibinfo{author}{\bibfnamefont{X.~C.}~\bibnamefont{Xie}}, \bibnamefont{and}
  \bibinfo{author}{\bibfnamefont{Q.}~\bibnamefont{Niu}},
  \bibinfo{journal}{Phys. Rev. Lett.} \textbf{\bibinfo{volume}{76}},
  \bibinfo{pages}{975} (\bibinfo{year}{1996}).

\bibitem[{\citenamefont{Xie et~al.}(1996)\citenamefont{Xie, Liu, Sundaram, and
  Niu}}]{XLSN96}
\bibinfo{author}{\bibfnamefont{X.~C.} \bibnamefont{Xie}},
  \bibinfo{author}{\bibfnamefont{D.~Z.} \bibnamefont{Liu}},
  \bibinfo{author}{\bibfnamefont{B.}~\bibnamefont{Sundaram}}, \bibnamefont{and}
  \bibinfo{author}{\bibfnamefont{Q.}~\bibnamefont{Niu}},
  \bibinfo{journal}{Phys. Rev. B} \textbf{\bibinfo{volume}{54}},
  \bibinfo{pages}{4966} (\bibinfo{year}{1996}).

\bibitem[{\citenamefont{Yang and Bhatt}(1996)}]{YB96}
\bibinfo{author}{\bibfnamefont{K.}~\bibnamefont{Yang}} \bibnamefont{and}
  \bibinfo{author}{\bibfnamefont{R.~N.} \bibnamefont{Bhatt}},
  \bibinfo{journal}{Phys. Rev. Lett.} \textbf{\bibinfo{volume}{76}},
  \bibinfo{pages}{1316} (\bibinfo{year}{1996}).

\bibitem[{\citenamefont{Sheng and Weng}(1997)}]{SW97}
\bibinfo{author}{\bibfnamefont{D.~N.} \bibnamefont{Sheng}} \bibnamefont{and}
  \bibinfo{author}{\bibfnamefont{Z.~Y.} \bibnamefont{Weng}},
  \bibinfo{journal}{Phys. Rev. Lett.} \textbf{\bibinfo{volume}{78}},
  \bibinfo{pages}{318} (\bibinfo{year}{1997}).

\bibitem[{\citenamefont{Sheng and Weng}(1998)}]{SW98}
\bibinfo{author}{\bibfnamefont{D.~N.} \bibnamefont{Sheng}} \bibnamefont{and}
  \bibinfo{author}{\bibfnamefont{Z.~Y.} \bibnamefont{Weng}},
  \bibinfo{journal}{Phys. Rev. Lett.} \textbf{\bibinfo{volume}{80}},
  \bibinfo{pages}{580} (\bibinfo{year}{1998}).

\bibitem[{\citenamefont{Potempa et~al.}(1998)\citenamefont{Potempa, B\"aker,
  and Schweitzer}}]{PBS98}
\bibinfo{author}{\bibfnamefont{H.}~\bibnamefont{Potempa}},
  \bibinfo{author}{\bibfnamefont{A.}~\bibnamefont{B\"aker}}, \bibnamefont{and}
  \bibinfo{author}{\bibfnamefont{L.}~\bibnamefont{Schweitzer}},
  \bibinfo{journal}{Physica B} \textbf{\bibinfo{volume}{256-258}},
  \bibinfo{pages}{591} (\bibinfo{year}{1998}).

\bibitem[{\citenamefont{Yang and Bhatt}(1999)}]{YB99}
\bibinfo{author}{\bibfnamefont{K.}~\bibnamefont{Yang}} \bibnamefont{and}
  \bibinfo{author}{\bibfnamefont{R.~N.} \bibnamefont{Bhatt}},
  \bibinfo{journal}{Phys. Rev. B} \textbf{\bibinfo{volume}{59}},
  \bibinfo{pages}{8144} (\bibinfo{year}{1999}).

\bibitem[{\citenamefont{Sheng and Weng}(2000)}]{SW00b}
\bibinfo{author}{\bibfnamefont{D.~N.} \bibnamefont{Sheng}} \bibnamefont{and}
  \bibinfo{author}{\bibfnamefont{Z.~Y.} \bibnamefont{Weng}},
  \bibinfo{journal}{Phys. Rev. B} \textbf{\bibinfo{volume}{62}},
  \bibinfo{pages}{15363} (\bibinfo{year}{2000}).

\bibitem[{\citenamefont{Morita et~al.}(2000)\citenamefont{Morita, Ishibashi,
  and Hatsugai}}]{MIH00}
\bibinfo{author}{\bibfnamefont{Y.}~\bibnamefont{Morita}},
  \bibinfo{author}{\bibfnamefont{K.}~\bibnamefont{Ishibashi}},
  \bibnamefont{and} \bibinfo{author}{\bibfnamefont{Y.}~\bibnamefont{Hatsugai}},
  \bibinfo{journal}{Phys. Rev. B} \textbf{\bibinfo{volume}{61}},
  \bibinfo{pages}{15952} (\bibinfo{year}{2000}).

\bibitem[{\citenamefont{Potempa and Schweitzer}(2001)}]{PS01}
\bibinfo{author}{\bibfnamefont{H.}~\bibnamefont{Potempa}} \bibnamefont{and}
  \bibinfo{author}{\bibfnamefont{L.}~\bibnamefont{Schweitzer}},
  \bibinfo{journal}{Physica B} \textbf{\bibinfo{volume}{298}},
  \bibinfo{pages}{52} (\bibinfo{year}{2001}).

\bibitem[{\citenamefont{Koschny et~al.}(2001)\citenamefont{Koschny, Potempa,
  and Schweitzer}}]{KPS01}
\bibinfo{author}{\bibfnamefont{T.}~\bibnamefont{Koschny}},
  \bibinfo{author}{\bibfnamefont{H.}~\bibnamefont{Potempa}}, \bibnamefont{and}
  \bibinfo{author}{\bibfnamefont{L.}~\bibnamefont{Schweitzer}},
  \bibinfo{journal}{Phys.\ Rev.\ Lett.} \textbf{\bibinfo{volume}{86}},
  \bibinfo{pages}{3863} (\bibinfo{year}{2001}).

\bibitem[{\citenamefont{Koschny and Schweitzer}(2002)}]{KS02}
\bibinfo{author}{\bibfnamefont{T.}~\bibnamefont{Koschny}} \bibnamefont{and}
  \bibinfo{author}{\bibfnamefont{L.}~\bibnamefont{Schweitzer}},
  \bibinfo{journal}{Physica E} \textbf{\bibinfo{volume}{12}},
  \bibinfo{pages}{654} (\bibinfo{year}{2002}).

\bibitem[{\citenamefont{Pereira and Schulz}(2002{\natexlab{a}})}]{PS02a}
\bibinfo{author}{\bibfnamefont{A.~L.~C.} \bibnamefont{Pereira}}
  \bibnamefont{and} \bibinfo{author}{\bibfnamefont{P.~A.}
  \bibnamefont{Schulz}}, \bibinfo{journal}{Physica E}
  \textbf{\bibinfo{volume}{12}}, \bibinfo{pages}{650}
  (\bibinfo{year}{2002}{\natexlab{a}}).

\bibitem[{\citenamefont{Huckestein}(2000)}]{Huc00}
\bibinfo{author}{\bibfnamefont{B.}~\bibnamefont{Huckestein}},
  \bibinfo{journal}{Phys. Rev. Lett.} \textbf{\bibinfo{volume}{84}},
  \bibinfo{pages}{3141} (\bibinfo{year}{2000}).

\bibitem[{\citenamefont{Pereira and Schulz}(2002{\natexlab{b}})}]{PS02b}
\bibinfo{author}{\bibfnamefont{A.~L.~C.} \bibnamefont{Pereira}}
  \bibnamefont{and} \bibinfo{author}{\bibfnamefont{P.~A.}
  \bibnamefont{Schulz}}, \bibinfo{journal}{Phys. Rev. B}
  \textbf{\bibinfo{volume}{66}}, \bibinfo{pages}{155323}
  (\bibinfo{year}{2002}{\natexlab{b}}).

\bibitem[{\citenamefont{MacKinnon and Kramer}(1981)}]{MK81}
\bibinfo{author}{\bibfnamefont{A.}~\bibnamefont{MacKinnon}} \bibnamefont{and}
  \bibinfo{author}{\bibfnamefont{B.}~\bibnamefont{Kramer}},
  \bibinfo{journal}{Phys. Rev. Lett.} \textbf{\bibinfo{volume}{47}},
  \bibinfo{pages}{1546} (\bibinfo{year}{1981}).

\bibitem[{\citenamefont{MacKinnon and Kramer}(1983)}]{MK83}
\bibinfo{author}{\bibfnamefont{A.}~\bibnamefont{MacKinnon}} \bibnamefont{and}
  \bibinfo{author}{\bibfnamefont{B.}~\bibnamefont{Kramer}},
  \bibinfo{journal}{Z. Phys. B} \textbf{\bibinfo{volume}{53}},
  \bibinfo{pages}{1} (\bibinfo{year}{1983}).

\bibitem[{\citenamefont{Shklovskii et~al.}(1993)\citenamefont{Shklovskii,
  Shapiro, Sears, Lambrianides, and Shore}}]{Sea93}
\bibinfo{author}{\bibfnamefont{B.~I.} \bibnamefont{Shklovskii}},
  \bibinfo{author}{\bibfnamefont{B.}~\bibnamefont{Shapiro}},
  \bibinfo{author}{\bibfnamefont{B.~R.} \bibnamefont{Sears}},
  \bibinfo{author}{\bibfnamefont{P.}~\bibnamefont{Lambrianides}},
  \bibnamefont{and} \bibinfo{author}{\bibfnamefont{H.~B.} \bibnamefont{Shore}},
  \bibinfo{journal}{Phys. Rev. B} \textbf{\bibinfo{volume}{47}},
  \bibinfo{pages}{11487} (\bibinfo{year}{1993}).

\bibitem[{\citenamefont{Zharekeshev and Kramer}(1995)}]{ZK95}
\bibinfo{author}{\bibfnamefont{I.~K.} \bibnamefont{Zharekeshev}}
  \bibnamefont{and} \bibinfo{author}{\bibfnamefont{B.}~\bibnamefont{Kramer}},
  \bibinfo{journal}{Phys. Rev. B} \textbf{\bibinfo{volume}{51}},
  \bibinfo{pages}{17239} (\bibinfo{year}{1995}).

\bibitem[{\citenamefont{Batsch et~al.}(1996)\citenamefont{Batsch, Schweitzer,
  {Kh. Zharekeshev}, and Kramer}}]{BSZK96}
\bibinfo{author}{\bibfnamefont{M.}~\bibnamefont{Batsch}},
  \bibinfo{author}{\bibfnamefont{L.}~\bibnamefont{Schweitzer}},
  \bibinfo{author}{\bibfnamefont{I.}~\bibnamefont{{Kh. Zharekeshev}}},
  \bibnamefont{and} \bibinfo{author}{\bibfnamefont{B.}~\bibnamefont{Kramer}},
  \bibinfo{journal}{Phys.\ Rev.\ Lett.} \textbf{\bibinfo{volume}{77}},
  \bibinfo{pages}{1552} (\bibinfo{year}{1996}).

\bibitem[{\citenamefont{Kantelhardt et~al.}(1998)\citenamefont{Kantelhardt,
  Bunde, and Schweitzer}}]{KBS98}
\bibinfo{author}{\bibfnamefont{J.~W.}~\bibnamefont{Kantelhardt}},
  \bibinfo{author}{\bibfnamefont{A.}~\bibnamefont{Bunde}}, \bibnamefont{and}
  \bibinfo{author}{\bibfnamefont{L.}~\bibnamefont{Schweitzer}},
  \bibinfo{journal}{Phys.\ Rev.\ Lett.} \textbf{\bibinfo{volume}{81}},
  \bibinfo{pages}{4907} (\bibinfo{year}{1998}).

\bibitem[{\citenamefont{Potempa and Schweitzer}(2002)}]{PS02}
\bibinfo{author}{\bibfnamefont{H.}~\bibnamefont{Potempa}} \bibnamefont{and}
  \bibinfo{author}{\bibfnamefont{L.}~\bibnamefont{Schweitzer}},
  \bibinfo{journal}{Phys.\ Rev.\ B} \textbf{\bibinfo{volume}{65}},
  \bibinfo{pages}{201105} (\bibinfo{year}{2002}).

\bibitem[{\citenamefont{Ando et~al.}(1982)\citenamefont{Ando, Fowler, and
  Stern}}]{AFS82}
\bibinfo{author}{\bibfnamefont{T.}~\bibnamefont{Ando}},
  \bibinfo{author}{\bibfnamefont{A.~B.} \bibnamefont{Fowler}},
  \bibnamefont{and} \bibinfo{author}{\bibfnamefont{F.}~\bibnamefont{Stern}},
  \bibinfo{journal}{Rev. Mod. Phys.} \textbf{\bibinfo{volume}{54}},
  \bibinfo{pages}{437} (\bibinfo{year}{1982}).

\bibitem[{\citenamefont{Ando}(1984)}]{And84}
\bibinfo{author}{\bibfnamefont{T.}~\bibnamefont{Ando}},
  \bibinfo{journal}{Journal of the Physical Society of Japan}
  \textbf{\bibinfo{volume}{53}}, \bibinfo{pages}{3126} (\bibinfo{year}{1984}).

\bibitem[{\citenamefont{Sheng et~al.}(2001)\citenamefont{Sheng, Weng, and
  Wen}}]{SWW01}
\bibinfo{author}{\bibfnamefont{D.~N.} \bibnamefont{Sheng}},
  \bibinfo{author}{\bibfnamefont{Z.~Y.} \bibnamefont{Weng}}, \bibnamefont{and}
  \bibinfo{author}{\bibfnamefont{X.~G.} \bibnamefont{Wen}},
  \bibinfo{journal}{Phys. Rev. B} \textbf{\bibinfo{volume}{64}},
  \bibinfo{pages}{165317} (\bibinfo{year}{2001}).


\end{thebibliography}

\end{document}